\documentclass{iau} 
\usepackage{graphicx}

\newcommand{\mnras}{Monthly Notices of the RAS}
\newcommand{\apj}{Astrophysical Journal}
\newcommand{\apjl}{Astrophysical Journal Letters}
\newcommand{\Ms}{{\ensuremath{{M}_{\odot} }}}

\title[FM 22.~~Ancient SNe in Frontier Fields] 
{Finding Ancient Supernovae at 5 $< z <$ 12 with Frontier Fields}

\author[Dan Whalen]   
{Daniel J. Whalen$^1$}

\affiliation{$^1$Zentrum f\"{u}r Astronomie, Institut f\"{u}r Theoretische Astrophysik, Universit\"{a}t 
Heidelberg, Albert-Ueberle-Str. 2, 69120 Heidelberg, Germany \\ email: {\tt dwhalen@uni-heidelberg.de}}

\pubyear{2015}
\volume{XXIX}  
\jname{The Frontier Fields: Transforming our understanding of cluster and galaxy evolution}
\editors{A.C. Editor, B.D. Editor \& C.E. Editor, eds.}
\begin{document}

\maketitle

\begin{abstract}

Supernovae are important probes of the properties of stars at high redshifts because they can be detected 
at early epochs and their masses can be inferred from their light curves. Direct detection of the first cosmic 
explosions in the universe will only be possible with JWST, WFIRST and the next generation of extremely 
large telescopes. But strong gravitational lensing by massive clusters, like those in the Frontier Fields, could 
reveal supernovae at slightly lower redshifts now by magnifying their flux by factors of 10 or more.  We find 
that Frontier Fields will likely discover dozens of core-collapse supernovae at 5 $ < z <$ 12.  Future surveys 
of cluster lenses similar in scope to Frontier Fields by JWST might find hundreds of these events out to $z 
\sim$ 15 - 17. Besides revealing the masses of early stars, these ancient supernovae could also constrain 
cosmic star formation rates in the era of first galaxy formation.

\keywords{early universe -- galaxies: high-redshift -- galaxies: clusters: general -- gravitational lensing -- 
large-scale structure of universe -- stars: early-type -- supernovae: general}
 
\end{abstract}

\firstsection 

\section{Introduction}

With the advent of the {\it James Webb Space Telescope} ({\it JWST}) and the next generation of 30 m class 
telescopes, it will soon be possible to detect SNe at the edge of the observable universe at $z \sim$ 10 - 20 
and use them to probe the earliest stellar populations (\cite[Whalen et al. 2008]{wet08c}; \cite[2013a]{wet12a
},\cite[b]{wet12b},\cite[c]{wet12c},\cite[d]{wet12d},\cite[e]{wet12e},\cite[f]{wet12f}; \cite[Whalen et al. 2014a]{
wet13d},\cite[b]{wet13e}; \cite[de Souza et al. 2013]{ds13};\cite[2014]{ds14}).  But SNe could be discovered in 
surveys at $z > $ 5 now.  In principle, strong lensing by massive galaxy clusters at $z \lesssim 1$ could boost 
flux from SNe by factors of 20 or more and allow them to be detected at $z \gtrsim$ 10 by HST (\cite[Whalen 
et al. 2013g]{clash}).  

We have computed the number of SNe expected to be found in the Frontier Fields (FF) survey at $z >$ 5. 
First, we construct magnification maps, $\mu$, as a function of redshift for each cluster in FF from its $\kappa$ 
and $\gamma$ maps, which are taken from \cite[Zitrin et al. 2015]{zitrin}.  Then we calculate the volume that is 
lensed to a given $\mu$ by a small patch on the map with that magnification, $dA(\mu,z)$, and sum the $dV$ to 
compute the total volume of space, $\mathrm{d}V(z,\mu)$, lensed to this $\mu$ at $z_i$ by the cluster.  We 
then convolve these lensed volumes with cosmic star formation rates (SFRs) from observations and simulations 
to calculate the number of SNe enclosed by the lensing volume $\mathrm{d}V(z,\mu)$ whose flux is boosted by 
a factor $\mu$ or more at redshift $z$ over the time d$t$ in the observer frame:
\begin{equation}  
\mathrm{d}N(z,\mu) \; = \; \mathrm{SFR}  \; (1 + z)  \; \frac{1  \; \mathrm{SN}}{100  \; \Ms}  
\; \mathrm{d}V(z,\mu) \frac{\mathrm{d}t}{1 + z}. \vspace{0.05in}
\label{eq:nsn}
\end{equation}  
To obtain the total number of SNe above a given redshift, $N(>z)$, we integrate $\mathrm{d}N(z,\mu)$ over all 
$z$ above that redshift and all $\mu$ above the minimum magnification needed to detect the event at each 
redshift.  

\begin{figure}
\begin{center}
\begin{tabular}{cc}
\includegraphics[width=0.45\textwidth]{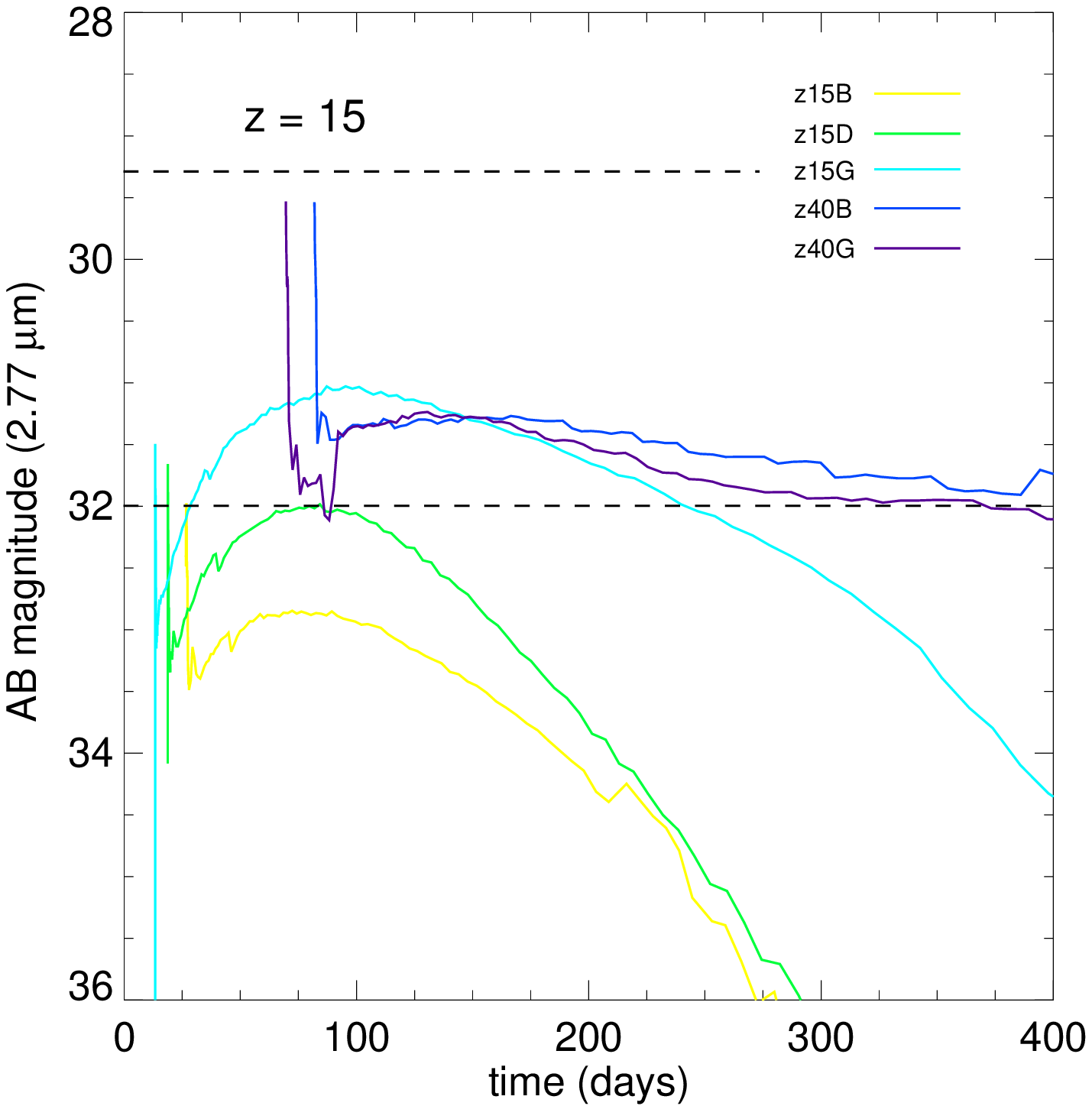} &
\includegraphics[width=0.45\textwidth]{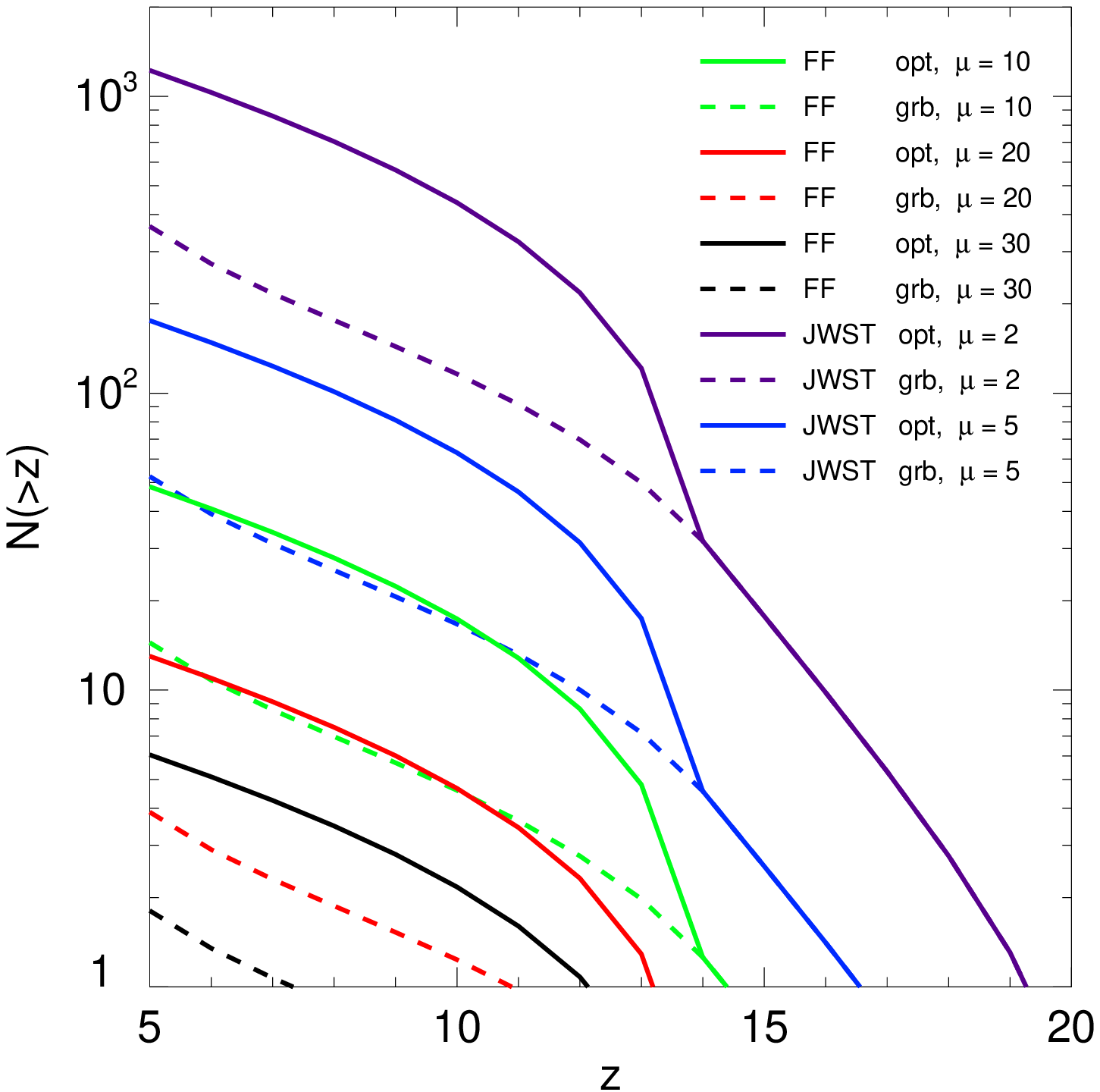} \\
\end{tabular}
\end{center}
\caption{Left panel: 15 - 40 \Ms\ Pop III CC SN NIR light curves at $z =$ 15.  The two dashed lines are FF and
JWST photometry limits at AB mag 29.3 and 32, respectively.  Right panel:  cumulative SN rates down to  
redshift $z$.} \vspace{0.1in}
\label{fig:LCs}
\end{figure}

Light curves for Pop III core-collapse SNe and detection rates are shown in Fig.~\ref{fig:LCs}.  FF could find 
dozens of SNe at 5 $ < z <$ 12, and future surveys of these clusters by JWST might find hundreds of SNe out 
to $z \sim$ 15 - 17.  Such detections will probe the properties of stars and their formation rates in the era of 
first galaxy formation.

\begin{thebibliography}{}

\bibitem[{{de Souza} {et~al.}(2013){de Souza}, {Ishida}, {Johnson}, {Whalen},
  \& {Mesinger}}]{ds13}
{de Souza}, R.~S., {Ishida}, E.~E.~O., {Johnson}, J.~L., {Whalen}, D.~J., \&
  {Mesinger}, A. 2013, \mnras, 436, 1555

\bibitem[{{de Souza} {et~al.}(2014){de Souza}, {Ishida}, {Whalen}, {Johnson},
  \& {Ferrara}}]{ds14}
{de Souza}, R.~S., {Ishida}, E.~E.~O., {Whalen}, D.~J., {Johnson}, J.~L., \&
  {Ferrara}, A. 2014, \mnras, 442, 1640

\bibitem[{{Whalen} {et~al.}(2008b){Whalen}, {Prochaska}, {Heger},
  \& {Tumlinson}}]{wet08c}
{Whalen}, D., {Prochaska}, J.~X., {Heger}, A., \& {Tumlinson}, J.
  2008b, \apj, 682, 1114
  
\bibitem[{{Whalen} {et~al.}(2013d){Whalen}, {Fryer}, {Holz},
  {Heger}, {Woosley}, {Stiavelli}, {Even}, \& {Frey}}]{wet12a}
{Whalen}, D.~J., {Fryer}, C.~L., {Holz}, D.~E., {Heger}, A., {Woosley}, S.~E.,
  {Stiavelli}, M., {Even}, W., \& {Frey}, L.~H. 2013d, \apjl, 762,
  L6

\bibitem[{{Whalen} {et~al.}(2013a){Whalen}, {Even}, {Frey},
  {Smidt}, {Johnson}, {Lovekin}, {Fryer}, {Stiavelli}, {Holz}, {Heger},
  {Woosley}, \& {Hungerford}}]{wet12b}
{Whalen}, D.~J., {Even}, W., {Frey}, L.~H., {Smidt}, J., {Johnson}, J.~L.,
  {Lovekin}, C.~C., {Fryer}, C.~L., {Stiavelli}, M., {Holz}, D.~E., {Heger},
  A., {Woosley}, S.~E., \& {Hungerford}, A.~L. 2013{a}, \apj, 777,
  110

\bibitem[{{Whalen} {et~al.}(2013e){Whalen}, {Joggerst}, {Fryer},
  {Stiavelli}, {Heger}, \& {Holz}}]{wet12c}
{Whalen}, D.~J., {Joggerst}, C.~C., {Fryer}, C.~L., {Stiavelli}, M., {Heger},
  A., \& {Holz}, D.~E. 2013e, \apj, 768, 95

\bibitem[{{Whalen} {et~al.}(2013c){Whalen}, {Even}, {Smidt},
  {Heger}, {Chen}, {Fryer}, {Stiavelli}, {Xu}, \& {Joggerst}}]{wet12d}
{Whalen}, D.~J., {Even}, W., {Smidt}, J., {Heger}, A., {Chen}, K.-J., {Fryer},
  C.~L., {Stiavelli}, M., {Xu}, H., \& {Joggerst}, C.~C. 2013c,
  \apj, 778, 17

\bibitem[{{Whalen} {et~al.}(2013b){Whalen}, {Even}, {Lovekin},
  {Fryer}, {Stiavelli}, {Roming}, {Cooke}, {Pritchard}, {Holz}, \&
  {Knight}}]{wet12e}
{Whalen}, D.~J., {Even}, W., {Lovekin}, C.~C., {Fryer}, C.~L., {Stiavelli}, M.,
  {Roming}, P.~W.~A., {Cooke}, J., {Pritchard}, T.~A., {Holz}, D.~E., \&
  {Knight}, C. 2013{b}, \apj, 768, 195

\bibitem[{{Whalen} {et~al.}(2013f){Whalen}, {Johnson}, {Smidt},
  {Heger}, {Even}, \& {Fryer}}]{wet12f}
{Whalen}, D.~J., {Johnson}, J.~L., {Smidt}, J., {Heger}, A., {Even}, W., \&
  {Fryer}, C.~L. 2013f, \apj, 777, 99

\bibitem[{{Whalen} {et~al.}(2013g){Whalen}, {Johnson}, {Smidt},
  {Meiksin}, {Heger}, {Even}, \& {Fryer}}]{wet12g}
{Whalen}, D.~J., {Johnson}, J.~L., {Smidt}, J., {Meiksin}, A., {Heger}, A.,
  {Even}, W., \& {Fryer}, C.~L. 2013g, \apj, 774, 64

\bibitem[{{Whalen} {et~al.}(2014a){Whalen}, {Smidt}, {Even},
  {Woosley}, {Heger}, {Stiavelli}, \& {Fryer}}]{wet13d}
{Whalen}, D.~J., {Smidt}, J., {Even}, W., {Woosley}, S.~E., {Heger}, A.,
  {Stiavelli}, M., \& {Fryer}, C.~L. 2014a, \apj, 781, 106

\bibitem[{{Whalen} {et~al.}(2014b){Whalen}, {Smidt}, {Heger},
  {Hirschi}, {Yusof}, {Even}, {Fryer}, {Stiavelli}, {Chen}, \&
  {Joggerst}}]{wet13e}
{Whalen}, D.~J., {Smidt}, J., {Heger}, A., {Hirschi}, R., {Yusof}, N., {Even},
  W., {Fryer}, C.~L., {Stiavelli}, M., {Chen}, K.-J., \& {Joggerst}, C.~C.
  2014b, \apj, 797, 9

\bibitem[Whalen et al. 2013]{clash} Whalen, D.~J., Smidt, 
J., Johnson, J.~L., et al.\ 2013, arXiv:1312.6330 

\bibitem[Zitrin et al. (2015)]{zitrin15} Zitrin, A., Fabris, A., 
Merten, J., et al.\ 2015, \apj, 801, 44 

\end{thebibliography}
\end{document}